\documentclass[runningheads]{llncs}
\usepackage{graphicx}
\usepackage{xcolor}
\usepackage{hyperref}
\usepackage{amsmath}
\usepackage{mathtools}
\usepackage{amssymb}
\usepackage[ruled,vlined]{algorithm2e}

\begin{document}
\raggedbottom
\title{Passive query-recovery attack against secure conjunctive keyword search schemes}
\author{Marco Dijkslag\inst{1} \and Marc Damie\inst{3} \and
  Florian Hahn\inst{1} \and Andreas Peter\inst{1,2}}
\authorrunning{M. Dijkslag et al.}
\titlerunning{Passive query-recovery attack against secure CKWS schemes}
%
\institute{University of Twente, Enschede, The Netherlands
  \email{m.dijkslag@alumnus.utwente.nl}, \email{\{f.w.hahn,a.peter\}@utwente.nl}\\ \and
  University of Oldenburg, Oldenburg, Germany\\ \email{andreas.peter@uni-oldenburg.de} \\ \and
  Univ. Lille, Inria, CNRS, Centrale Lille, UMR 9189 - CRIStAL, Lille, France\\
  \email{marc.damie@inria.fr}} \setlength{\intextsep}{16pt}
\maketitle              
\newcommand{\kw}{\textit{kw}}
\newcommand{\ckw}{\textit{ckw}}
\newcommand{\td}{\textit{td}}
\newcommand{\known}{\textit{known}}
\newcommand{\real}{\textit{real}}
\newcommand{\similar}{\textit{similar}}

\newcommand{\Dreal}{$\mathcal{D}_{\real}$}
\newcommand{\Dsimilar}{$\mathcal{D}_{\similar}$}
\newcommand{\Dpknown}{$\mathcal{D}_{p-\known}$}

\newcommand{\Cckw}{$C_{\ckw}$}
\newcommand{\Ctd}{$C_{\td}$}
\newcommand{\Ckw}{$C_{\kw}$}

\newcommand{\W}{$\mathcal{W}$}
\newcommand{\Wreal}{$\mathcal{W}_{\real}$}
\newcommand{\Wknown}{$\mathcal{W}_{\known}$}
\newcommand{\Wsimilar}{$\mathcal{W}_{\similar}$}

\newcommand{\K}[1]{$\mathcal{K}_{#1}$}
\newcommand{\Kreal}{$\mathcal{K}_{\real}$}
\newcommand{\Kknown}{$\mathcal{K}_{\known}$}
\newcommand{\Ksimilar}{$\mathcal{K}_{\similar}$}

\begin{abstract}
  While storing documents on the cloud can be attractive, the question remains whether cloud 
  providers can be trusted with storing private documents. Even if trusted, data breaches are 
  ubiquitous. To prevent information leakage one can store documents encrypted. If encrypted 
  under traditional schemes, one loses the ability to perform simple operations over the 
  documents, such as searching through them. Searchable encryption schemes 
  were proposed allowing some search functionality while documents remain encrypted.
  Orthogonally, research is done to find attacks that exploit search and access pattern 
  leakage that most efficient schemes have. One type of such an attack is the ability to 
  recover plaintext queries. Passive query-recovery attacks on single-keyword search 
  schemes have been proposed in literature, however, conjunctive keyword search has not 
  been considered, although keyword searches with two or three keywords appear more 
  frequently in online searches.
  
  We introduce a generic extension strategy for existing passive query-recovery attacks 
  against single-keyword search schemes and explore its applicability for the attack 
  presented by Damie et al. (USENIX Security '21). While the original attack achieves up 
  to a recovery rate of $85\%$ against single-keyword search schemes for an attacker 
  without exact background knowledge, our experiments show that the generic extension 
  to conjunctive queries comes with a significant performance decrease achieving 
  recovery rates of at most $32\%$. Assuming a stronger attacker with partial 
  knowledge of the indexed document set boosts the recovery rate to $85\%$ for 
  conjunctive keyword queries with two keywords and achieves similar recovery 
  rates as previous attacks by Cash et al. (CCS '15) and Islam et al. 
  (NDSS '12) in the same setting for single-keyword search schemes. 
  
  \keywords{Searchable encryption \and conjunctive keyword search \and
    passive query-recovery attack.}
\end{abstract}
\section{Introduction}

With increasing number of enterprises storing their documents in the cloud the question arises 
how to cope with 
storing sensitive documents on the cloud without the cloud provider learning information about
the stored documents or information being leaked when a data breach occurs. One solution for 
this problem would be to encrypt the documents to 
hide its contents to the cloud provider. However, this prevents users from using the 
(often available) computational resources cloud providers offer, since searching through 
the documents is no longer possible without first downloading and decrypting it. 

Searchable symmetric encryption schemes can be a solution to this problem that offer 
constructions for search functionalities over encrypted documents. The first practical 
solution towards searchable encryption has been proposed by Song et al. 
\cite{song2000practical}. 
Proposers of searchable encryption schemes need to find a trade-off in efficiency, security, 
and functionality. With this trade-off in terms of security comes information leakage such as 
possible \textit{search pattern} leakage (revealing which queries concerned the same underlying, 
but unknown, keyword) and \textit{access pattern} leakage (revealing the identifiers of all  
documents matching the search query). Most of the efficient searchable encryption schemes 
that allow for keyword search leak information in the access pattern for efficiency. 

Searchable encryption is an active line of research for finding efficient schemes that 
allow for search in encrypted documents with well-defined security in terms of a leakage 
function. Orthogonally, research is performed on finding attacks against proposed searchable 
encryption schemes. One such type of attack is a query-recovery attack, i.e. the ability for 
an adversary to recover the plaintexts from performed queries. In general two kinds of 
query-recovery attacks exist: (1) a \textit{passive} attack where an adversary only has 
access to the information leaked by a scheme and (2) an \textit{active} attack in which 
an adversary is able to inject tailored documents into the to-be-searched dataset. 

\textit{Active} query-recovery attacks on conjunctive keyword search do exist 
\cite{zhang2016all,poddar2020practical} which are described as an extension on the proposed 
single-keyword search attack. Currently, all existing \textit{passive} query-recovery 
attacks against searchable symmetric encryption that allow for keyword searches only focuses 
on single-keyword search schemes. However, these attacks do not reflect a realistic scenario,
since single-keyword searches are limited and statistics show that the number 
of keywords used by people online in the US peaks at two keywords \cite{clement_2020}. Also, 
three keyword searches are still more frequent than searches for a single keyword. 
The frequency of searches using seven or more keywords becomes negligible.

Note that the recovery of conjunctive keyword 
queries is more difficult with respect to the recovery of single-keyword queries using 
similar vocabulary sizes. This difficulty stems from the fact that the space for keyword 
conjunctions is combinatorial in the number of conjunction terms compared to 
single-keywords, therefore an attacker needs to consider more possible candidates of 
keyword conjunctions for each observed query.

In this work, we explore a passive query-recovery 
attack against secure conjunctive keyword search (CKWS) schemes. 
We propose a generic extension strategy for query-recovery attacks against 
single-keyword search to recover conjunctive queries using the same attack. Our 
extension strategy is based on the use of trapdoors created from a keyword-conjunction 
set as a generalization of trapdoors created from single-keywords. Replacing
keywords with keyword conjunction sets. Our attack is static and does also 
work on forward and backward private schemes (\cite{patranabis2020forward}).

We introduce an adaptation of the query-recovery attack proposed by Damie et al. 
\cite{queryvolution2020} to achieve keyword conjunction recovery. We explore the 
applicability of the attack in two setups: (1) a \textit{similar-documents} attack, where 
the attacker only has access to a set of documents that is similar, but otherwise different, 
from the indexed documents and (2) a \textit{known-documents} attack, where the attacker has 
(partial) knowledge of the indexed documents. In both 
setups it is assumed the attacker knows the keyword conjunctions for a small set of queries 
a priori. We experimentally show that our attack can work for a relatively small vocabulary
size ($500$) in an attack setup allowing only conjunctive keyword search using $2$ keywords.
However, we show that in an attack setup using similar-documents the attack performs poorly 
unless many known queries are assumed to be part of the attacker's knowledge. 
Furthermore, we demonstrate limitations of our generic extension posed by the 
combinatorial complexity increase for larger conjunctions.

\section{Related work}\label{section:related-work}
Most attacks against searchable symmetric encryption that have been described in the 
literature are query-recovery attacks.
Islam et al. \cite{islam2012access} were the first to propose a passive query-recovery 
attack in which they are exploiting the \textit{access pattern} leakage, i.e. leaked 
document identifiers from observed queries. In their attack, the adversary needs to know 
all the documents indexed on the server to be successful. 
They introduced the idea of computing (word-word and trapdoor-trapdoor) co-occurrences to 
attack SSE. This idea being reused by other the passive attacks. 
The attack works by finding the closest mapping between the word-word co-occurrence 
matrix and trapdoor-trapdoor co-occurrence matrix 
in which they use meta heuristic simulated annealing. Also, the attack requires a number 
of known queries to work, i.e. trapdoors from which the attacker knows the underlying 
plaintext value.

Cash et al. \cite{cash2015leakage} proposed another passive query-recovery attack. 
Their attack
first exploits that keywords with high frequency have unique keyword document counts to 
initialize their set of known queries. Then for keywords that do not have a unique keyword 
document occurrence count they construct a co-occurrence matrix of their known documents and 
observed queries, similar to Islam et al. They try to recover more queries by constructing 
for every unknown query their candidate set (i.e. keywords having the same document 
occurrence count) and remove candidates from the set that do not have the same 
co-occurrence with a known query in the known queries set. If after iterating over every 
known query only one candidate is left, the last candidate 
is appended to the known queries set. This process is repeated for all unknown queries 
until the set of known queries stops increasing.

Both \cite{islam2012access,cash2015leakage} rely on the attacker knowing a large part of
the indexed documents, where the count attack performs better than the attack by Islam et al. 
However, their query recovery rate roughly only increases when the attacker knows at least
80\% of the indexed documents.

The query-recovery attack proposed by Pouliot et al. \cite{pouliot2016shadow} uses 
weighted graph matching where the attacker needs to find mapping of keyword graph $G$ and 
trapdoor graph $H$. The attack achieves recovery rates above $90\%$
when the attacker knows the entire set of indexed documents, but fails as similar-documents 
attack unless having a smaller set of documents and vocabulary size. Also, the runtime of the 
attack increases rapidly, where for a vocabulary size of $500$ the attack runs in less than one 
hour, whereas it takes more than $16$ hours for a vocabulary size of $1000$. 
The attack in \cite{islam2012access} has a runtime of a maximum of $14$ hours, whereas 
attacks from \cite{cash2015leakage,queryvolution2020} run in seconds.

Ning et al. \cite{ning2018passive} introduced a query-recovery attack that works when 
the attacker knows a percentage of the indexed documents. Keywords and trapdoors are 
represented as a binary string where the $i$-th bit is $1$ if the keyword 
(resp. trapdoor) occurs in document $i$.
Recovery is done by converting the bit strings to integers, where it is considered that 
a keyword corresponds to a trapdoor if they have the same integer value.

The proposed attack outperforms the attack 
by Cash et al. \cite{cash2015leakage}, where in their scenario \cite{cash2015leakage}
achieves a recovery rate of roughly $28\%$ and their proposed attack around $56\%$ when 
the attacker knows $80\%$ of the indexed documents. However, they do not report a
recovery rate for an attacker having knowledge of more than $80\%$ of the 
indexed documents.

Blackstone et al. \cite{blackstone2019revisiting} proposed a "sub-graph" attack requiring 
much less known documents to be successful and also works on co-occurrence hiding schemes. 
Their experiments show that an attacker only needs to know $20\%$ of the indexed documents 
to succeed in her attack.. 

In \cite{queryvolution2020}, Damie et al. proposed their refined score attack that works 
in a setting where the attacker only knows a similar, but otherwise different and 
non-indexed, set of documents for query-recovery. 
A mathematical formalization of the similarity is proposed in their paper.
In \cite{cash2015leakage} they showed that both the attack proposed by
Islam et al. \cite{islam2012access} and their proposed count attack do not work using similar 
documents. In \cite{queryvolution2020}, the query-recovery attack uses similar techniques 
as used by \cite{islam2012access,cash2015leakage}, i.e. constructing co-occurrence matrices 
from the document set known by the attacker and a trapdoor-trapdoor co-occurrence matrix from 
the assumed access pattern leakage. By starting with a few known 
(keyword, trapdoor)-pairs their attack iteratively recovers queries where previous 
recovered queries with high confidence scores are added to the set of known queries. 
Using this approach their attack reaches recovery rates around $85\%$. 

\textbf{Other types of attacks.}
Zhang et al. \cite{zhang2016all} proposed an effective active document injection attack 
to recover keywords. Furthermore, they proposed an extension of their attack to a 
conjunctive keyword search setting which was experimentally verified for queries 
with $3$ keywords.

In \cite{poddar2020practical}, Poddar et al. proposed several attacks that 
uses volume pattern as auxiliary information in combination with the attacker's ability 
to replay queries and inject documents. Moreover, they also gave an extension of their 
attack for queries with conjunctive keywords which is based on the extension from 
\cite{zhang2016all} using a document injection approach. 

Liu et al. \cite{liu2014search} proposed a query-recovery attack which makes use of the
search pattern leakage as auxiliary information. In particular, they exploit the query 
frequency. However, they simulated their queries by applying Gaussian noise to keyword 
search frequency from Google Trends\footnote{\url{https://trends.google.com/trends}} because 
of the lack of a query dataset. The attacker has access to the original frequencies.

Another attack introduced by Oya and Kerschbaum \cite{oya2021hiding} combines both 
volume information derived from the access pattern leakage and query frequency information
derived from the search pattern leakage as auxiliary information.

\textbf{Conjunctive keyword search schemes.}
Passive query-recovery attacks against single-keyword search schemes 
already work for some conjunctive keyword search schemes where the server
performs search for each individual keyword in a query independently and returns the
intersection of document identifiers of each single-keyword search, i.e. leaking the 
\textit{full} access pattern for each individual keyword in the conjunction. However, 
these attacks cannot be applied on conjunctive keyword search schemes with less or 
\textit{common} access pattern leakage, where \textit{common} refers to the scheme only 
leaking the document identifiers for the documents containing all keywords from a 
conjunctive keyword query. Hence, in this work we explore one extension strategy for 
conjunctive keywords that can be applied to most passive query-recovery attacks against 
single-keyword search using only common access pattern leakage.

\cite{poon2015efficient,sun2015catch}
both proposed such a conjunctive keyword search scheme that returns the intersection of 
document identifiers for each individual keyword in a conjunctive keyword query, thus 
leaking the \textit{full} access pattern. However, we would 
like to emphasize that in this scenario only an \textit{honest-but-curious} server that 
is able to observe the result set for each intermediate keyword can be 
considered an attacker, since an \textit{eavesdropper} on the communication channel 
would not be able to observe the document identifiers for each intermediate single-keyword 
search. 
Furthermore, it should be noted that both schemes also offer more functionality 
than conjunctive keyword search alone. Where 
\cite{poon2015efficient} allows for phrase searches and \cite{sun2015catch} offers result 
set verifiability and index updatability.

Other proposed conjunctive keyword search schemes exist 
(\cite{golle2004secure,cash2013highly,jho2013symmetric,fairouz2016symmetric,zhang2018privacy,lai2018result,wu2019vbtree,hu2019forward,poddar2020practical}).
However, all of them leak at least the common access pattern, where 
\cite{cash2013highly,hu2019forward,wang2019toward} have more than common access pattern 
leakage. To the best of our knowledge there do not exist efficient 
conjunctive keyword search schemes that have no access pattern leakage.

\section{Preliminaries}
\begin{table*}[h!]
  \caption{Notation}
  \label{tab:notations}
  \begin{tabular}{lll}
    \hline
    \textbf{Notation} & \textbf{Meaning}                                                       & \textbf{Size notation}                           \\
    \hline
    $\mathcal{Q}$     & Set of observed trapdoors by the adversary                             & $l$                                              \\
    $R_Q$             & Document identifiers for each observed $\td \in \mathcal{Q}$           & $l$                                              \\
    \textit{KnownQ}   & Known ($\td$, $\ckw$)-pairs by the adversary                           & $k$                                              \\
    $\ckw_q$          & 
    \begin{tabular}{@{}l@{}} Set of distinct keywords used in \\
      a conjunctive keyword query $q$\end{tabular}
    
                      & $d$                                                                                                                       \\
    \hline
    \Cckw             & \begin{tabular}{@{}l@{}} $\ckw$-$\ckw$ co-occurrence matrix created \\
                          from \Dsimilar $ $ or \Dpknown\end{tabular}
                      & 
    \begin{tabular}{@{}l@{}}
      $m_{\similar} \times m_{\similar}$ \\or $m_{\known} \times m_{\known}$
    \end{tabular} \\
    \Ctd              & $\td$-$\td$ co-occurrence matrix created from $R_Q$                    & $l \times l$                                     \\
    \hline
    \Dreal            & Real (indexed) document set                                            & $n_{\real}$                                      \\
    \Dsimilar         & Similar document set                                                   & $n_{\similar}$                                   \\
    \Dpknown          & $p$-Known document set ($0 < p \leq 1$)                                & $n_{\known}$
    ($= p \cdot n_{\real}$)                                                                                                                       \\
    \hline
    \Wreal            & Vocabulary of keywords extracted from \Dreal
                      & $v_{\real}$                                                                                                               \\
    \Wsimilar         & Vocabulary of keywords extracted from 
    \Dsimilar         & $v_{\similar}$                                                                                                            \\
    \Wknown           & Vocabulary of keywords extracted 
    from \Dpknown     & $v_{\known}$                                                                                                              \\
    \hline
    \Kreal            & 
    \begin{tabular}{@{}l@{}} Set containing possible conjunctions of keyword \\
      combinations generated from \Wreal\end{tabular}
                      & $m_{\real} = {v_{\real} \choose d}$                                                                                       \\
    \Ksimilar         & 
    \begin{tabular}{@{}l@{}} Set containing possible conjunctions of keyword \\
      combinations generated from \Wsimilar\end{tabular}
                      & $m_{\similar} = {v_{\similar} \choose d}$                                                                                 \\
    \Kknown           & 
    \begin{tabular}{@{}l@{}} Set containing possible conjunctions of keyword \\
      combinations generated from \Wknown\end{tabular}
                      & $m_{\known} = {v_{\known} \choose d}$                                                                                     \\
    \hline
  \end{tabular}
\end{table*}

We first introduce some notations that are used throughout this work.
Let document set $\mathcal{D}$ consist of documents $\{D_1, ..., D_n\}$. 
Let keyword set $\mathcal{W}$ consist
of keywords $\{w_1, ..., w_m\}$. Document $D_i$ consists of keywords that form a subset 
of keyword set $\mathcal{W}$. Let $id(D_i) = i$ return the identifier for document $D_i$. 
We denote $x \in D_i$ if keyword $x$ ($\in \mathcal{W}$) occurs in document $D_i$. 
A summary of all notations and their meaning used throughout this work is given in 
Table \ref{tab:notations}.

\subsection{Searchable symmetric encryption}

A searchable encryption scheme allows a user to search in encrypted documents and is often 
described in a client-server setting. The client can search through encrypted documents stored 
on the server, without the server learning information about the plaintext documents.
Often a searchable encryption scheme can be divided in four algorithms:

\begin{itemize}
  \item $\mathsf{KeyGen}(1^k)$: takes security parameter $k$ and outputs a secret key $K$.
  \item $\mathsf{BuildIndex}(K, \mathcal{D})$: takes document set $\mathcal{D}$ and secret
        key $K$ and produces an (inverted) index $I$.
  \item $\mathsf{Trapdoor}(K, q)$: takes query $q$ and secret key $K$ and outputs a trapdoor
        $td_q$.
  \item $\mathsf{Search}(I, \td_q)$: takes trapdoor $\td_q$ and index $I$ and outputs the
        documents that match with query $q$.
\end{itemize}

In single-keyword search schemes $q$ corresponds to a keyword $w$, 
whereas in conjunctive keyword search schemes $q$ would correspond to a query for 
documents containing $d$ keywords, i.e., the conjunction of keywords 
$w_1 \wedge$ ... $\wedge w_d$ of keywords $w_1$, ..., $w_d$. Then, $\td_q$ would 
correspond to the conjunction of $d$ keywords.

\subsection{Considered conjunctive keyword search model}\label{subsection:conjunctive-model}

We assume a fixed number of keywords ($d$) that are allowed to be searched for in a 
conjunctive keyword search. For instance if $d = 2$, only trapdoors with $2$ distinct 
keywords are allowed. We denote such a fixed-$d$ scheme as 
\textit{secure $d$-conjunctive keyword search scheme}.

For simplicity, we assume a fixed number 
of $d$ distinct keywords, however one could consider $d$ as a maximum number of keywords 
in the conjunctive search by reusing the same keyword for non-used keyword entries 
in the conjunction. For instance, when $d = 2$, $kw \wedge kw$ for the same keyword 
$kw$ would be equivalent to a single-keyword search for $kw$.

We consider $\ckw$ to be the set of $d$ different keywords that are used to construct
a trapdoor ($\td_{\ckw}$).
For instance, if we consider a conjunctive keyword search scheme that allows search 
for $d = 3$ conjunctive keywords, we would create a keyword set $\ckw$ for 
every possible combination of $3$ keywords, where $\ckw_1 = \{kw_1, kw_2, kw_3\}$.
\footnote{Note: $d = 1$ refers to a single-keyword search scheme.}

First, in the $\mathsf{BuildIndex}$ algorithm, the client encrypts every document in the document 
set locally. Then creates an encrypted index of the document set (locally). 
Given a trapdoor $td_{ckw}$, the server can find the documents containing keywords in $ckw$
using such a created index. The encrypted document set and index are then uploaded by the client
to the server. 

Although in literature different methods for constructing such an index were 
proposed, here we do not fix which index is used. We only require the model to have 
at least \textit{common} access pattern leakage, where \textit{common} refers to the 
scheme only leaking the document identifiers for the documents containing all 
keywords in a conjunctive keyword query. All conjunctive search 
schemes described in Section \ref{section:related-work} leak at least the 
common access pattern.

The client can search documents by constructing trapdoors. The client constructs a 
trapdoor by picking $d$ keywords she wants to search for. In our model, she 
constructs a trapdoor using the function $\td_q = \mathsf{Trapdoor}(K, \ckw_i = \{kw_1, ...,\\ kw_d\})$,
for the keywords she wants to search for. By sending the trapdoor $\td_q$ to the server, 
the server responds with a set of 
document identifiers $R_{\td_q}$ for documents that contain all keywords in $\ckw_i$.

\subsection{Attacker model}

Like in \cite{queryvolution2020}, we consider two types of passive attackers which both can
observe trapdoors sent by a user and its response including the document identifiers. 
The first type of attacker is an \textit{honest-but-curious} server. The server is 
considered to be an honest entity meaning it follows the protocol. Hence, it always 
returns the correct result for each query. However, such curious server tries to learn as 
much information as possible using the scheme leakage. Secondly, we consider an 
\textit{eavesdropper} that is able to observe pairs of trapdoor and document identifiers
from the communication channel between client and server as an attacker. 

For both attackers an $observation_i$ is a tuple $(\td_q, R_{\td_q})$ considering 
conjunctive keyword queries where trapdoor $\td$ corresponds to $d$ conjunctive keywords.

\subsection{Attacker knowledge}

It is assumed the attacker knows the number of keywords $d$ that are allowed to 
construct trapdoors. Moreover, it can be assumed that an 
\textit{honest-but-curious} attacker knows the byte size of the stored documents 
and the number of documents stored (e.g. from the index). However, an \textit{eavesdropper}
does not. In that case we make 
use of the proposed formula by \cite{queryvolution2020} that approximates the number of 
documents stored on the server ($n_{\real}$) derived from the attacker's 
knowledge.

We consider two types of attack setups, i.e. a \textit{similar-documents} attack setup 
where the attacker has access to a set of similar documents (as formalized in 
\cite{queryvolution2020}) and a \textit{known-documents} attack setup where the attacker has 
(partial) knowledge of the documents stored on the server. 

\textbf{Similar-documents attack.} In our similar-documents 
attack we assume the attacker has a document set \Dsimilar $ $ that is 
$\epsilon$-similar to the real indexed document set \Dreal. However, we assume 
$\epsilon$-similarity (as formalized in \cite{queryvolution2020}) over the possible keyword 
conjunctions rather than keywords, where smaller $\epsilon$ means more similar. Also, 
\Dsimilar $ $ $ \cap $ \Dreal $ $ $ = \emptyset$, thus do not have overlapping documents. 

\textbf{Known-documents attack.}
Like in \cite{islam2012access,cash2015leakage}, 
for our known-documents attack setup we 
assume that the attacker has a $p$-known document set \Dpknown, where $0 < p \leq 1$
defines the known-documents rate. Meaning, the attacker knows a fraction $p$ of the 
real indexed document set \Dreal $ $ stored on the server.

It should be noted that a similar-documents attack can be considered more 
realistic than a known-documents attack as discussed by Damie et al. \cite{queryvolution2020}. 
Since a known-documents attack will most likely only be possible on a data breach, whereas 
documents that are only similar to the actual indexed documents maybe even publicly available. 
Moreover, the user could remove the leaked documents that are used in a known-documents 
attack from the index.

The assumption that the attacker knows (a subset of) the documents stored on the server
is rather strong, but is based on what is done in previous work \cite{islam2012access,cash2015leakage}.

\section{CKWS-adapted refined score attack}

In this section we describe our conjunctive keyword search (CKWS) adaptation of the 
\textit{refined score attack}. Our adaptation builds upon the \textit{score attacks}
that were introduced by Damie et al. 
\cite{queryvolution2020}. We have chosen to use their query-recovery attack 
against single-keyword search schemes, since it is, to the best of our knowledge, 
the most accurate similar-documents attack that has been described yet.
Furthermore, the matching algorithm used in their attack 
only has a runtime of $20$ seconds while considering a vocabulary size of 
$4000$ keywords. Since the space of possible queries increases combinatorial, we have to 
consider many possible keyword conjunctions and thus faster runtimes is desired. 
Moreover, their attack can use either known documents or similar documents as adversary's 
knowledge. We describe how one can transform their query-recovery attack to an attack 
on conjunctive keyword search schemes, i.e. considering the (abstract) secure 
$d$-conjunctive keyword search scheme described in Section 
\ref{subsection:conjunctive-model}, using similar terminology as in \cite{queryvolution2020}.

In addition, the code for the score attacks has been made publicly available online by 
Damie et al. This allowed us to verify their results first before adapting 
it to our conjunctive keyword setting.

\subsection{Score attacks}

Damie et al. \cite{queryvolution2020} first propose the \textit{score attack} based on 
the idea of ranking potential keyword-trapdoor mappings according to a \textit{score} function. 
To run the score attack an attacker calculates the word-word co-occurrence matrix from its 
auxiliary document set and constructs a trapdoor-trapdoor co-occurrence matrix from observed 
queries and their result sets. Assuming some known queries, the attacker removes the columns 
from both matrices that do not occur in their set of known queries (i.e. word-trapdoor pairs) 
to obtain so-called sub-matrices. Then for every (observed) trapdoor, it goes through all 
possible keywords extracted from the auxiliary document set and returns the keyword for 
which their score function is maximized. 

Secondly, their proposed \textit{refined score attack} builds upon previously described score 
attack. Instead of returning a prediction for all trapdoors, they define a \textit{certainty}
function for each prediction and only keep the $RefSpeed$ best predictions according to this 
certainty function. These predictions are then 
added to the set of known queries and the attacker recomputes the co-occurrence sub-matrices. 
This procedure is repeated until there are no predictions left to make, i.e. no unknown queries 
left.

\subsection{Generic extension}

In short, our generic extension proposes to replace single keywords with 
keyword conjunction sets. The extension consists of five steps, highlighted by the 
next five subsections to adapt a passive query-recovery 
attack against single-keyword search to conjunctive keyword search, i.e. attacks that try 
to find a mapping between co-occurrences of keywords and trapdoors to recover queries.
We describe our extension in a similar-documents attack setup using 
$\mathcal{D}_{\similar}$, but the same steps can be taken in a known-documents attack setup 
using $\mathcal{D}_{p-\known}$ as the attacker's auxiliary document set.

\textbf{Extract vocabulary.} First, the attacker extracts keywords from the 
set of documents $\mathcal{D}_{\similar}$ to vocabulary $\mathcal{W}_{\similar}$.
As in query-recovery attacks on single-keyword search 
\cite{islam2012access,cash2015leakage,queryvolution2020}, we also assume 
that the keyword extraction method used by the attacker is the same as the one used 
by the user when she created the encrypted index. 

\textbf{Construct set of possible keyword conjunctions.} The attacker creates 
the set of all possible keyword conjunctions $\mathcal{K}_{\similar} = \{\ckw_i \in
  \mathcal{P}(\mathcal{W}_{\similar}) \bigm| |\ckw_i| = d\}$, where $m_{\similar} =
  |\mathcal{K}_{\similar}| = {v_{\similar} \choose d}$ and $\mathcal{P}(X)$ denotes 
the power set of set $X$. 

\textbf{Compute co-occurrence matrix for keyword conjunctions.} From \\
$\mathcal{D}_{\similar}$ and derived keyword conjunctions set $\mathcal{K}_{\similar}$
the attacker creates the $m_{\similar} \times m_{\similar}$ matrix $ID_{\similar}$. Here 
$ID_{\similar}[i,j] = 1$ if the $i$-th document in $\mathcal{D}_{\similar}$ contains 
the keywords that are in keyword conjunction $\ckw_j$ and is otherwise $0$. Then the 
attacker computes the $\ckw$-$\ckw$ co-occurrence matrix 
$C_{\ckw} = ID^T_{\similar} \cdot ID_{\similar} \cdot \frac{1}{n_{\similar}}$.
\footnote{$A^T$ denotes the transpose of matrix A.}

\textbf{Compute the trapdoor-trapdoor co-occurrence matrix.} We define 
$\mathcal{Q} = \{\td_1, ..., \td_l\}$ to be the set of observed queries by the attacker 
containing trapdoors that have been queried by the user. These trapdoors were created 
by the user from keyword conjunctions in $\mathcal{K}_{\real} = \{\ckw_i \in
  \mathcal{P}(\mathcal{W}_{\real}) \bigm| |\ckw_i| = d\}$.
Let $R_{\td} = \{id(D) | (\ckw \in \mathcal{K}_{\real}) \wedge 
  (\td = \mathsf{Trapdoor}(K, \ckw)) \wedge (D \in \mathcal{D}_{\real}) \wedge \forall_{kw_t
    \in \ckw} (kw_t \in D) \}$
be the set of document identifiers that were observed by the attacker for trapdoor $\td$.
Then we define the set of document identifiers 
$DocumentIDs = \bigcup_{\td \in \mathcal{Q}} R_{\td}$ of size $s$, where $s \leq n_{\real}$.
Similar to the construction of the matrix $ID_{\similar}$, we construct 
$s \times l$ trapdoor-document matrix $ID_{\real}$, 
where $ID_{\real}[i,j] = 1$ if $i$-th document identifier occurs in $R_{\td_j}$
(and $td_j$ refers to $j$-th trapdoor from $\mathcal{Q}$). Otherwise, 
$ID_{\real}[i,j] = 0$. Then trapdoor-trapdoor co-occurrence matrix 
$C_{\td} = ID^T_{\real} \cdot ID_{\real} \cdot \frac{1}{n_{\real}}$.

\textbf{Apply attack.} The last step is to apply a passive query-recovery 
attack using the set of keyword conjunctions and the co-occurrence matrices.

\subsection{Transform key steps of refined score attack}

As in \cite{islam2012access,cash2015leakage,queryvolution2020}, our attack also requires
the attacker to have knowledge of a set of known queries. However, our set of known 
queries is slightly different because of the keyword conjunctions. In 
a similar-documents attack setup our set of known queries $KnownQ =
  \{ (\ckw_i, \td_{\known}) | (\ckw_i \in \mathcal{K}_{\similar} \cap 
  \mathcal{K}_{\real}) \land (\td_{\known} \in \mathcal{Q}) \land 
  (\td_{\known} = \mathsf{Trapdoor}(K, \ckw_i) \}$. For our known-documents attack setup, 
$KnownQ$ is similarly defined by replacing $\mathcal{K}_{\similar}$ with 
$\mathcal{K}_{\known}$.

We recall key steps in the score attack w.r.t. the projection of 
the keyword-keyword co-occurrence and trapdoor-trapdoor co-occurrence matrix to 
sub-matrices using the set of known queries. These steps are important because they are 
different for our CKWS-adapted refined score attack. In short, the projection is done by only 
keeping the columns of known queries in $C_{\ckw}$ and $C_{\td}$.

Our goal is to generate sub-matrices $C^s_{\ckw}$ and $C^s_{\td}$ from $C_{\ckw}$ and 
$C_{\td}$ respectively. We describe the projection step for $C_{\ckw}$ using 
$\mathcal{K}_{\similar}$, but the same holds for $\mathcal{K}_{\known}$. Recall that 
$\mathcal{K}_{\similar} = \{\ckw_1, ..., \ckw_{m_{\similar}}\}$. 

We define $pos(\ckw)$, which returns the position of 
$\ckw \in \mathcal{K}_{similar}$. That is, $pos(\ckw_i) = i$. Similarly, 
$pos(\addcontentsline{}{}{}td)$ returns the 
position of $\td$ in $\mathcal{Q} = \{\td_1, ..., \td_l\}$. 

Let $C_{\ckw} = \begin{pmatrix}\hdots, \vec c_i, \hdots\end{pmatrix}_{i \in [m_{\similar}]}$ be the 
$m_{\similar} \times m_{\similar}$ co-occurrence matrix, where the column vector 
$\vec c_i$ denotes its $i$-th column. Then the $m_{\similar} \times k$ sub-matrix 
$C^s_{\ckw}$ $=$ $\begin{pmatrix} \hdots, \vec c_{pos(\ckw_j)}, \hdots \end{pmatrix}_{(\ckw_j, \td_j) \in KnownQ}$,
where $\vec c_{pos(\ckw_j)}$ is the $pos(\ckw_j)$-th column vector of $C_{\ckw}$.

Let $C_{\td} = \begin{pmatrix} \hdots, \vec u_{i} , \hdots
  \end{pmatrix}_{i \in [l]}$ be the $l \times l$ trapdoor-trapdoor co-occurrence matrix, where 
the column vector $\vec u_i$ denotes its $i$-th column. Then $l \times k$ sub-matrix 
$C^s_{\td}$ can be constructed as follows:
$C^s_{\td} = \begin{pmatrix} \hdots, \vec u_{pos(\td_j)}, \hdots \end{pmatrix}_{(\ckw_j, \td_j) \in KnownQ}$,
where $u_{pos(\td_j)}$ is the $pos(\td_j)$-th column vector of $C_{\td}$.

Superscript $s$ emphasizes that $C^s_{\ckw}$ and $C^s_{\td}$
are sub-matrices of $C_{\ckw}$ and $C_{\td}$ respectively. Also, we 
denote $C^s_{\ckw}[\ckw_i]$ to be the $i$-th row vector for 
keyword conjunction set $\ckw_i$ and $C^s_{\td}[\td_j]$ to be the $j$-th row vector for 
trapdoor $\td_j$, where $|C^s_{\ckw}[\ckw_i]| = |C^s_{\td}[\td_j]| = k$.

Additionally, we revise the \textit{scoring algorithm} for which the score is higher 
if a trapdoor corresponds 
to a certain keyword conjunction, i.e. the distance between two vectors 
$C^s_{\td}[\td_j]$ and $C^s_{\ckw}[\ckw_i]$ is small. Using keyword conjunctions the 
score function is defined as: $Score(\td_j, \ckw_i) = -ln(||C^s_{\ckw}[\ckw_i] - C^s_{\td}[\td_j]||)$,
for all $\ckw_i \in \mathcal{K}_{\similar}$ (or $\mathcal{K}_{\known}$) and all 
$\td_j \in \mathcal{Q}$, where $ln(\cdot)$ is the natural log and $||\cdot||$
is a vector-norm (e.g. L2 norm).

\subsection{Revised algorithm}

We substitute $C^s_{kw}$ for $C^s_{\ckw}$ in \cite{queryvolution2020}
to transform the refined score attack to the 
\textit{CKWS-adapted refined score attack}. 
Algorithm \ref{algorithm:ckwsadapted} contains its pseudocode, where a step 
is highlighted blue if it is different from the \textit{refined score attack}
proposed by Damie et al. \cite{queryvolution2020}. Note that this algorithm is 
described using $\mathcal{K}_{\similar}$, but also works for 
$\mathcal{K}_{\known}$ as input.

\begin{algorithm}[h!]
  \SetAlgoLined
  \KwIn{${\color{blue} \mathcal{K}_{\similar}}$, ${\color{blue} C^s_{\ckw}}$, $\mathcal{Q}$, 
  $C^s_{\td}$, $KnownQ$, $RefSpeed$}
  \KwResult{List of keyword conjunctions as predictions for trapdoors with certainty}
  $final\_pred \gets []$\;
  $unknownQ \gets \mathcal{Q}$\;
  \While{$unknownQ \neq \emptyset$}{
  \tcp{Set remaining unknown queries.}
  $unknownQ \gets \{ td: (td \in \mathcal{Q}) \wedge (\nexists {\color{blue} \ckw}
    \in {\color{blue} \mathcal{K}_{\similar}}: (td, {\color{blue} \ckw}) \in KnownQ)\}$\;
  $temp\_pred \gets []$\;
  \BlankLine
  \tcp{Propose a prediction for each unknown query.}
  \ForAll{$\td \in unknownQ$}{
  $cand \gets []$\;
  \ForAll{${\color{blue} \ckw} \in {\color{blue} \mathcal{K}_{\similar}}$}{
  $s \gets -ln(||{\color{blue} C^s_{\ckw}[\ckw]} - C^s_{\td}[\td]||)$\;
  Append \{“kw”: ${\color{blue} \ckw}$, “score”: $s$ \} to $cand$\;
  }
  Sort $cand$ in descending order according to the score\;
  $certainty \gets score(cand[0]) - score(cand[1])$\;
  Append $(\td, cand[0], certainty)$ to $temp\_pred$\;
  }
  \BlankLine
  \tcp{Stop refining or keep refining.}
  \eIf{$|unknownQ| < RefSpeed$}{
  $final\_pred \gets KnownQ \cup temp\_pred$\;
  $unknownQ \gets \emptyset$\;}
  {
  Add $RefSpeed$ most certain predictions $temp\_pred$ to $KnownQ$\;
  Add the columns corresponding to the new known queries to ${\color{blue} C^s_{\ckw}}$ and $C^s_{\td}$
  }
  }
  \Return{$final\_pred$}
  \caption{CKWS-adapted refined score attack.}
  \label{algorithm:ckwsadapted}
\end{algorithm}

One iteration of the algorithm can be defined by the three key phases. First 
remove known queries from the observed queries set $\mathcal{Q}$. Secondly, find the 
best scoring keyword conjunction candidate for each unknown query and 
compute the certainty of this candidate. Using keyword conjunctions the certainty of 
a keyword conjunction candidate $\ckw_i$ for trapdoor $\td$ is defined by:
$Certainty(\td, \ckw_i)$ $=$ $Score(\td, \ckw_i)$ $-$ $\max_{j \neq i} Score(\td, \ckw_j)$

Using this definition the certainty of a correct match of keyword conjunction with a 
trapdoor is higher when the score of the match is much higher than all other possible 
candidate scores.

The algorithm defines a notion of refinement speed ($RefSpeed$) which defines the 
number of most certain predictions that will be added each iteration of the algorithm
to the set of known queries. Which describes the third and last key step of an 
iteration, i.e. adding the most certain predictions to the known queries and recompute
sub-matrices $C^s_{\ckw}$ and $C^s_{\td}$. Thereafter, either start a new iteration or 
stop the algorithm if the number of unknown queries is less than $RefSpeed$.

\subsection{Complexity}\label{section:complexity}

As in \cite{queryvolution2020}, a higher refinement speed will result in a faster runtime,
but less accurate predictions. However, due to our use of keyword conjunctions the number 
of candidates for a trapdoor increases for larger $d$. Therefore, the runtime of the 
\textit{CKWS-adapted refined score attack} grows combinatorial. The time complexity 
of the attack is given by $\mathcal{O}(f(v) + g(v))$, where 
$f(v) = \frac{v!}{d!(v - d)!} \cdot (d - 1)$ corresponds to the time complexity of the 
generic extension, where we assume multiplying two vectors takes constant time. Further, 
$g(v) = \frac{|\mathcal{Q}|}{RefSpeed} \cdot |\mathcal{Q}| \cdot
  \frac{v!}{d!(v - d)!} \cdot k$ is the time complexity of the attack. For both $f$ and $g$, 
input $v$ is either $v_{\similar}$ or $v_{\known}$ depending on the attack setup.

Besides the increase in runtime, having $d > 1$ also the space complexity of the 
algorithm increases faster relative to the vocabulary size. Since co-occurrence matrix 
$C_{\ckw}$ in the similar-documents attack setup is $m_{\similar} \times m_{\similar}$, in 
terms of vocabulary size is $\frac{v_{\similar}!}{d!(v_{\similar} - d)!} \times
  \frac{v_{\similar}!}{d!(v_{\similar} - d)!}$ thus increasing faster with larger 
$v_{\similar}$.

This increase in time and space complexity led us to first further optimize the
revised algorithm for our implementations. Moreover, we use a GPU to decrease runtimes 
through computing expensive matrix operations on it.

\section{Experiments}\label{section:experiments}

\subsection{Setup}\label{section:experiments:setup}

\textbf{Documents.}
As described previously, in our experiments we simulate our 
attack using the publicly available
Enron email document set introduced by Klimt \& Yang \cite{klimt2004introducing}. 
We chose this document set since this one is also used in most attack papers requiring a set 
of documents. Similarly, we constructed the same corpus of emails from the folder 
$\_sent\_mail$ which results in a set of $30109$ documents.

\textbf{Keyword extraction.}
We extract keywords from solely the contents of the emails in 
the dataset, i.e. we do not consider email addresses or email subjects to be part of the 
document set. For keyword extraction we use the Porter Stemmer algorithm 
\cite{porter1980algorithm} to 
obtain stemmed words, moreover we remove stop words in the English language like 'the' or 
'a'. Using this method results in a total of $62976$ unique keywords in our entire 
considered document set.

\textbf{Number of keywords in conjunction.}
Throughout our experiments we fix $d$, i.e. 
the number of keywords allowed in one conjunction, 
to either $1$, $2$ or $3$. This means that no mixture of number of keywords is 
allowed in search. For instance, when the $d = 3$ only queries with $3$ distinct 
keywords are allowed, i.e. queries that contain either $1$ or $2$ keywords are not allowed.

\textbf{Testing environment.}
We implemented the attack on an Ubuntu $20.04$ server with Intel Xeon $20$-core processor 
($64$ bits, $2.2$ GHz), $512$ GB of memory, and NVIDIA Tesla P100 GPU 
($16$GB). We used Python $3.7$ and the Tensorflow library \cite{abadi2016tensorflow}
to accelerate matrix operations on a GPU.\footnote{Our code is available at
  \url{https://github.com/marcowindt/passive-ckws-attack}}

\textbf{Limitations.}
Running experiments with larger vocabulary sizes requires a lot of 
memory, since a vocabulary size of $150$ and $d = 2$ means a 
document-keyword-conjunction matrix size of $18065 \times 11175$ (already $1.5$ GiB) and 
a maximum co-occurrence matrix size of $11175 \times 11175$ ($0.9$ GiB) which
both have to fit in the memory of the GPU for fast calculations. Therefore, having similar
vocabulary sizes as used in the score attack is unrealistic in our generic 
extension strategy setting without having sufficient resources. However, we propose an 
extrapolation strategy to have approximate results for larger vocabularies.

\subsection{Results}\label{section:experiments:results}

In our experiments where similar-documents are used as the attacker's knowledge, we use the 
same ratio in similar (40\%) and real (60\%) documents as in \cite{queryvolution2020}. 
Similar to \cite{islam2012access,cash2015leakage,queryvolution2020}, we define the 
accuracy to be the number of correct predictions divided by the 
number of unknown queries excluding the initial known queries, i.e. 
the $accuracy = \frac{|CorrectPredictions(unknownQ)|}{|\mathcal{Q}| - |KnownQ|}$.

If not specified otherwise, each accuracy result corresponds to the average accuracy over $50$
experiments. Also, the vocabulary used in experiments is always created from the most 
frequently occurring keywords in the document set. From this vocabulary the keyword 
conjunctions set is generated. In each experiment it is assumed the attacker has observed 
$15\%$ of queries that can be performed by the user, i.e. $|\mathcal{Q}| = 0.15 \cdot m_{\real}$, 
where queries are sampled u.a.r. from $\mathcal{K}_{\real}$ to construct trapdoors.

\begin{figure}[h!]
  \centering
  \begin{minipage}{0.49\textwidth}
    \centering
    \includegraphics[width=\textwidth]{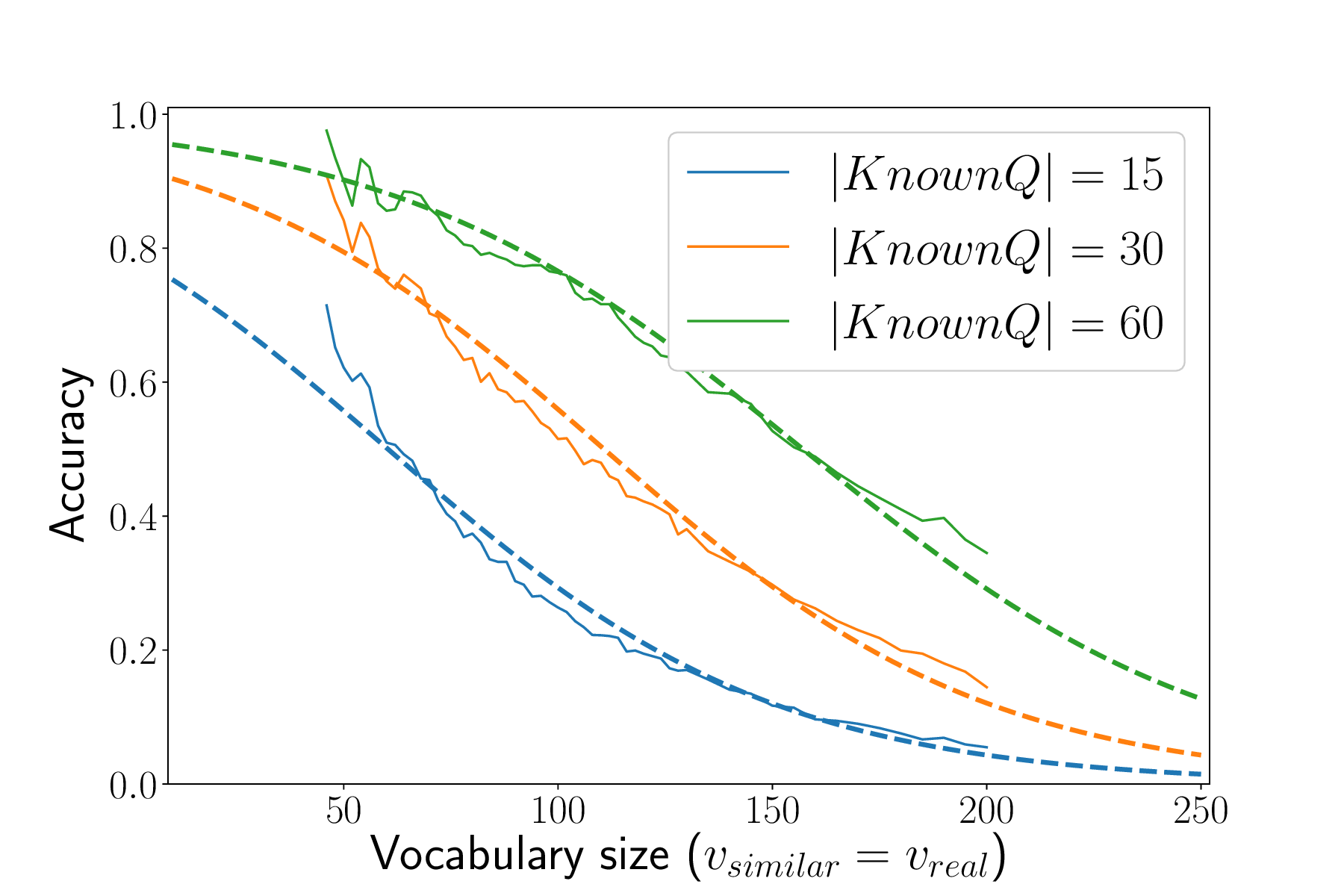}
    \caption{Score attack using similar-documents for varying vocabulary sizes and initially known queries
      with $d = 2$, $|\mathcal{D}_{\real}| = 18K$, $|\mathcal{D}_{\similar}| = 12K$, $|\mathcal{Q}| = 0.15 \cdot m_{\real}$.}
    \label{fig:basescore}
  \end{minipage}\hfill
  \begin{minipage}{0.49\textwidth}
    \centering
    \includegraphics[width=\textwidth]{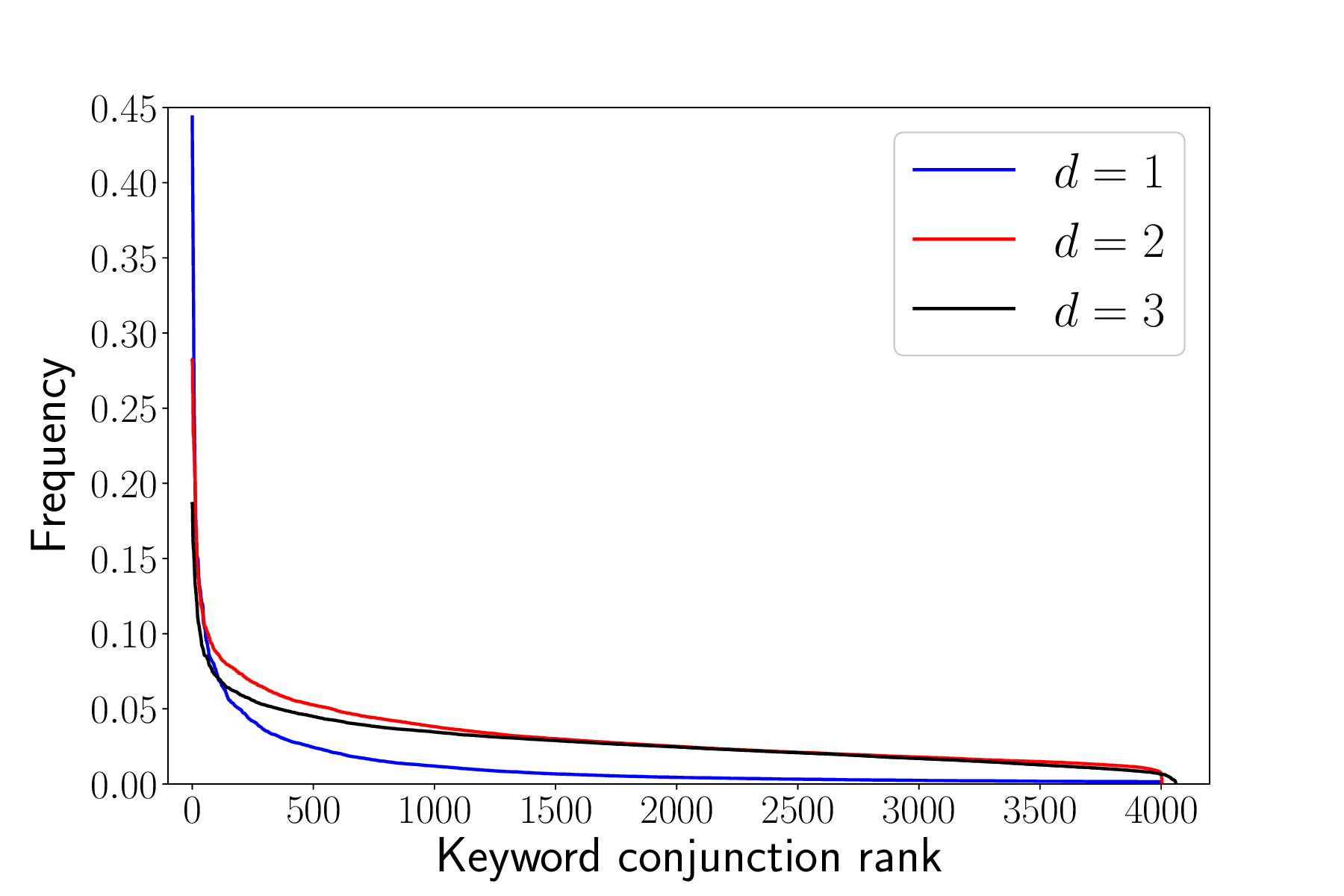}
    \caption{Frequency of keyword conjunctions ordered from most frequent to least frequent
      occurring keyword conjunction in $\mathcal{D}_{\similar}$.}
    \label{fig:frequencies}
  \end{minipage}
\end{figure}

\textbf{Result extrapolation.}

Figure \ref{fig:basescore} shows the accuracy of the score attack from 
\cite{queryvolution2020} where the attacker has access to similar-documents for varying 
vocabulary size and $d = 2$. We show these results to highlight that we can 
extrapolate the accuracy of the attack in a similar-documents setting closely, where 
the extrapolation is depicted by the dashed line and measured results are the solid line. 
We obtain this extrapolation by first transforming the accuracies using the 
$\mathsf{logit}$\footnote{$\mathsf{logit}(x) = \mathsf{log}(\frac{x}{1 - x})$} function. 
Using this transformation, we obtain a space in which we seem to have a linear 
relationship such that $\mathsf{logit}(acc) = b \cdot v_{\similar} + a$. We then perform 
a linear regression to obtain these coefficients using our experimental results. Lastly, 
we use the inverse $\mathsf{logit}$ function to transform it back to the original scale. 
We make use of this extrapolation where running experiments becomes infeasible (i.e. 
experiments with $d = 2$ and $v_{\real} > 500$) to extrapolate the accuracy for larger 
vocabulary sizes.

In our linear regressions, we do not provide the coefficient of determination 
$R^2$ and the $p$-value since they are based on the assumption that results are 
independent which is not true in our experiments as they all use the same document 
set. Hence, these values should not be used to evaluate the quality of the model 
even if they are high (e.g. $R^2 \approx 0.95$ in Figure 1) but the linear regression 
is still valid.
Although there may exist more precise extrapolation techniques, our intention is 
to have a simple yet realistic approximation of the accuracy for larger vocabularies 
for the sake of our discussion.

\textbf{Frequency of keyword conjunctions.}
Figure \ref{fig:frequencies} shows the frequency of a keyword conjunction occurring in 
$\mathcal{D}_{\similar}$ for $d \in \{1, 2, 3\}$, where keyword conjunction rank is lowest for the 
most frequent keyword conjunction. We observe the behavior of using keyword conjunctions 
instead of a single-keyword, i.e. the frequency of the most frequent keyword conjunction becomes 
smaller with higher $d$ and the frequency of the least frequent keyword conjunction reaches almost 
zero. This is to be expected, since the larger vocabulary size the higher the probability that 
certain keywords from a keyword conjunction do not appear in any document together, i.e. considering 
the vocabulary is generated with the most frequent keywords first. Note however, that the frequency
for rank between $200$ and $3600$ part is higher for $d = 2$ relative to $d = 1$, which is due to 
the fact that obtaining $4000$ keyword conjunctions requires a smaller vocabulary size of $90$ for 
$d = 2$, and it is still the case that the most frequent keywords occur together. Nevertheless, the 
same does not hold for $d = 3$ relative to $d = 2$, where we actually observe a decrease in keyword 
conjunction frequency. Here it already is the case that the most frequent keywords used to create a 
keyword conjunction of $3$ keywords do not have to necessarily occur together in a document.

\begin{figure}[h!]
  \centering
  \begin{minipage}{0.49\textwidth}
    \centering
    \includegraphics[width=0.99\textwidth]{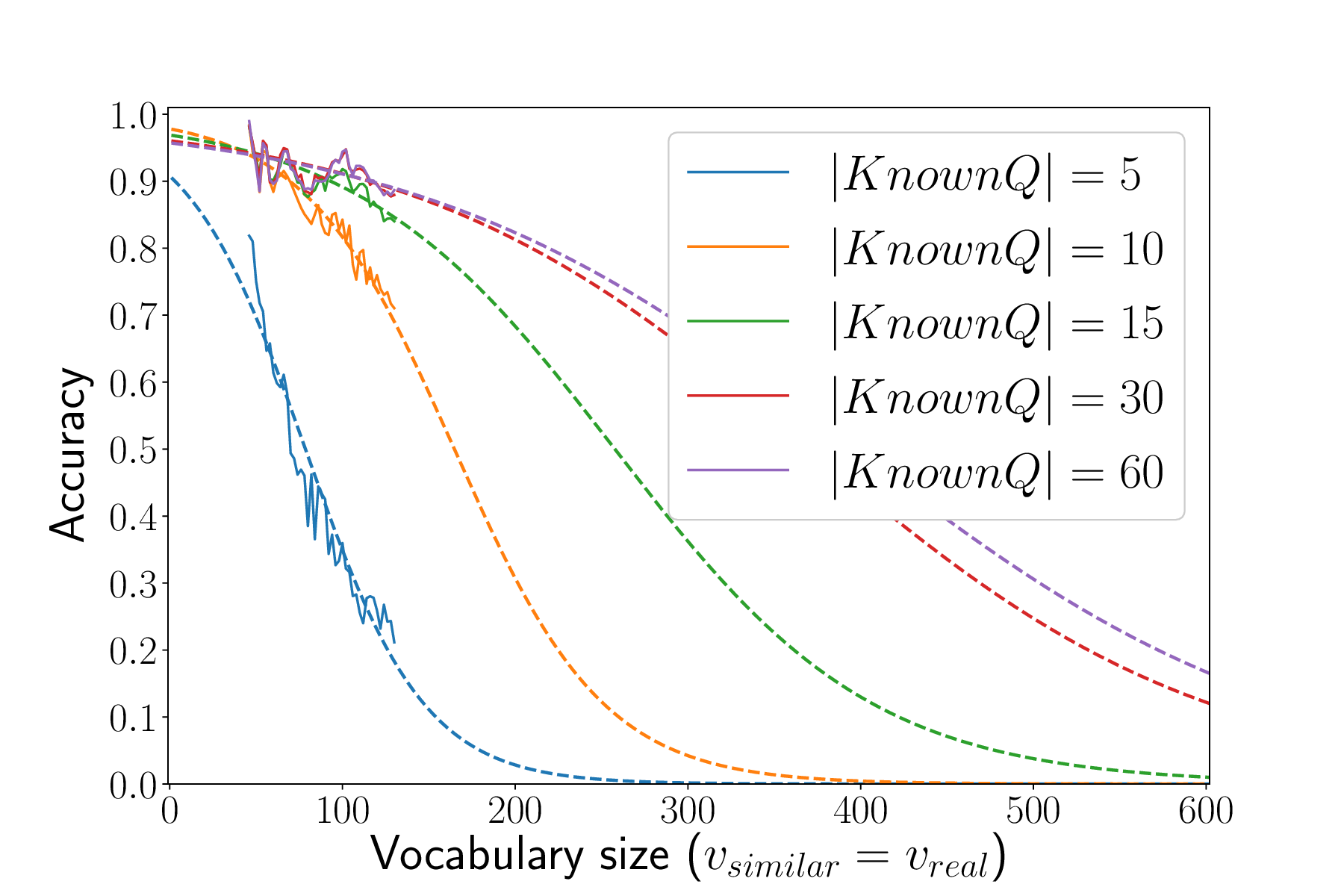}
    \caption{Accuracy plot of the CKWS-adapted refined score attack with
      $d = 2$ extrapolated and varying vocabulary size. With $|\mathcal{D}_{\real}| = 18K$, 
      $|\mathcal{D}_{\similar}| = 12K$, $|\mathcal{Q}| = 0.15 \cdot m_{\real}$.}
    \label{fig:similarrefined}
  \end{minipage}\hfill
  \begin{minipage}{0.49\textwidth}
    \centering
    \includegraphics[width=0.99\textwidth]{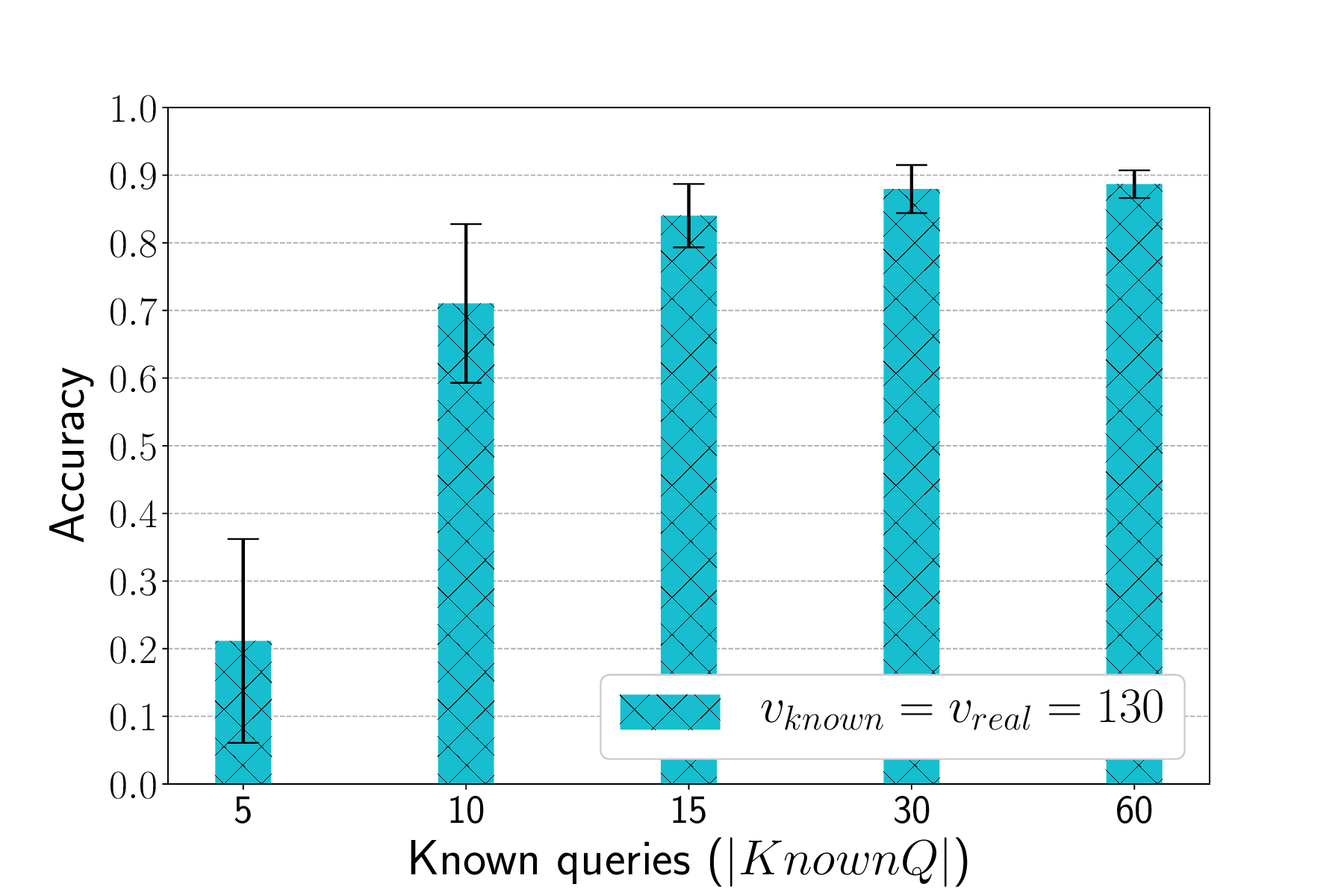}
    \caption{Accuracy plot of the CKWS-adapted refined score attack with
      $d = 2$ and varying number of known queries. With $|\mathcal{D}_{real}| = 18K$, 
      $|\mathcal{D}_{similar}| = 12K$, $|\mathcal{Q}| = 0.15 \cdot m_{real}$.}
    \label{fig:variance}
  \end{minipage}
\end{figure}

\textbf{CKWS-adapted refined score attack using similar-documents.}\label{subsubsection:notworking}
Figure \ref{fig:similarrefined} shows the accuracy of the CKWS-adapted refined score attack using 
similar-documents with 
$d = 2$ and varying vocabulary size. Also, the plot shows an extrapolation of the accuracies 
for vocabulary sizes larger than $130$ (and smaller than $50$). From the extrapolation 
of the accuracies for varying vocabulary sizes we clearly see a rapid decrease in accuracy with 
larger vocabulary sizes. We conclude that, when we consider the results with $30$ known queries 
we can still reach a reasonable recovery rate above $50\%$ for vocabulary size $300$ to 
$400$ keywords. However, the results are far from the single-keyword 
search set up presented in \cite{queryvolution2020} achieving up to $85\%$ recovery rate for 
vocabulary size of $1000$.

In \cite{queryvolution2020}, they discussed how the 'quality' of a known query influences the 
accuracy. A known query is more qualitative if the underlying keyword occurs more 
frequently. We remind that in the CKWS-adapted setting, it is a way to reduce the number of 
known queries needed. A lower rank of a keyword conjunction in Figure \ref{fig:frequencies}
the query for the keyword-conjunction is considered more qualitative. 

Figure \ref{fig:variance} shows the accuracy of the CKWS-adapted refined score attack using 
similar-documents with $d = 2$ and varying number of known queries. The plot shows that the 
standard deviation of the accuracy, assuming $5$ or $10$ known queries, 
is relatively high compared to the standard deviation for $15$, $30$, or $60$ known queries. 
For $5$ known queries the standard deviation is $0.15$, which is at least $3$ times higher 
than the standard deviation for $15$ known queries ($\approx 0.05$). 
The accuracy increases and standard deviation decreases with a higher number of known queries, since 
it becomes more likely to pick more qualitative queries (u.a.r.). This also explains 
why we observe this noisy behavior of the accuracy in the plot.

\begin{figure}[h!]
  \centering
  \begin{minipage}{0.49\textwidth}
    \centering
    \includegraphics[width=0.99\textwidth]{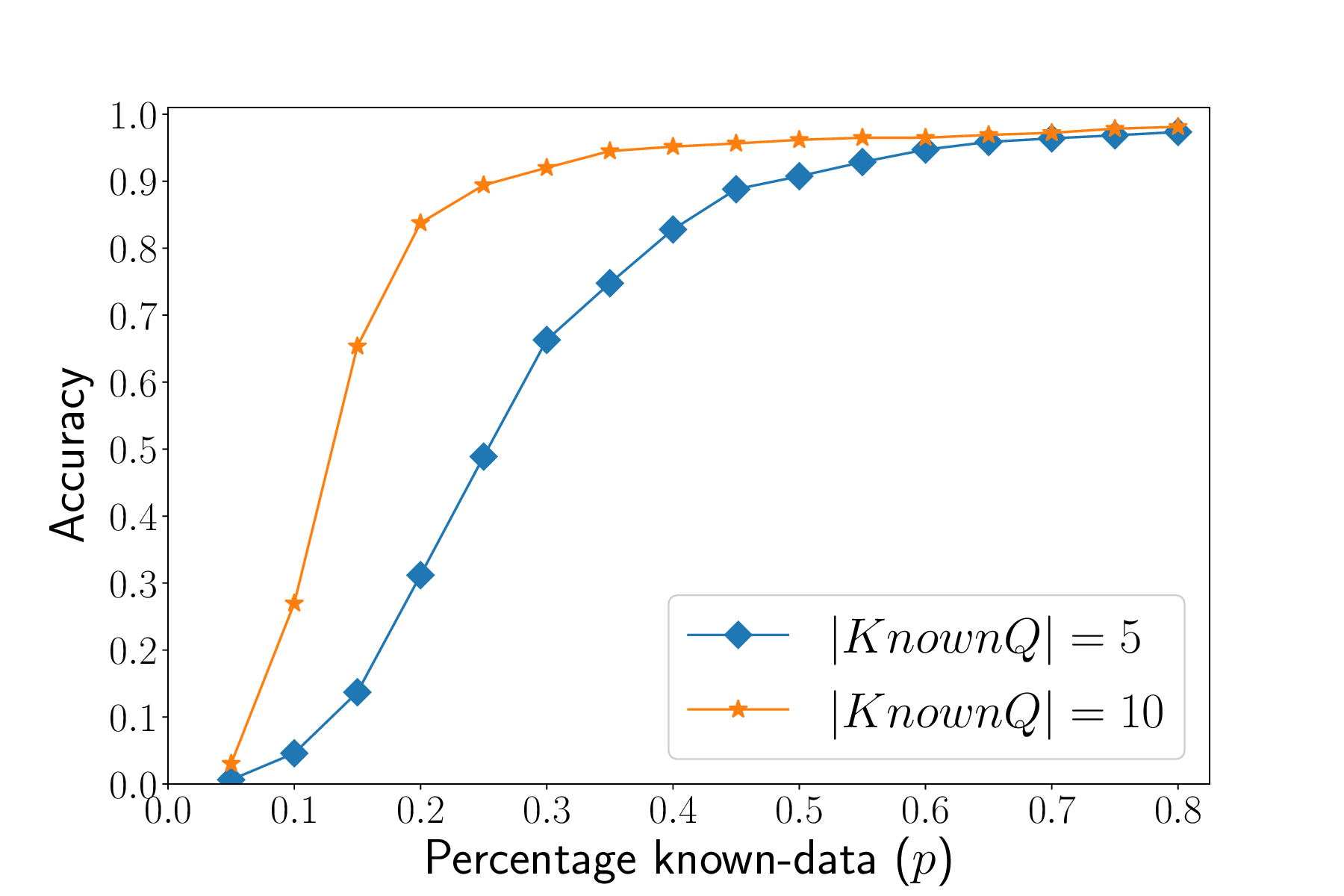}
    \caption{Accuracy of the CKWS-adapted refined score attack using known-data for varying known-data
      rates $p$ with $d = 2$ and $v_{known} = v_{real} = 130$.}
    \label{fig:varyingpknown}
  \end{minipage}\hfill
  \begin{minipage}{0.49\textwidth}
    \centering
    \includegraphics[width=0.99\textwidth]{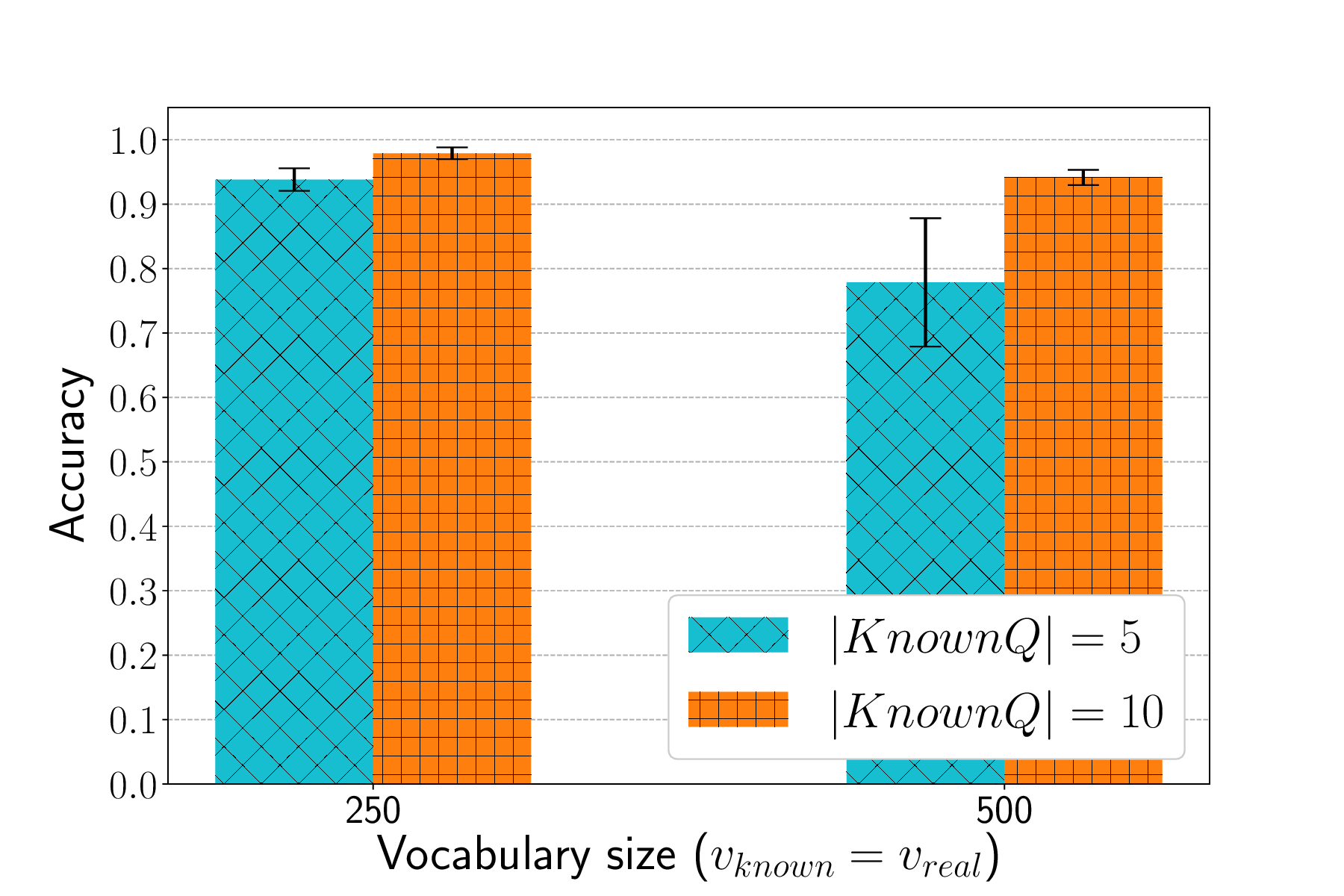}
    \caption{Accuracy of the extended refined score known-documents attack
      with $d = 2$ and $p = 0.7$, i.e. $n_{\known} = p \cdot n_{\real}$.}
    \label{fig:knownbars}
  \end{minipage}
\end{figure}

\textbf{CKWS-adapted refined score attack using $p$-known-documents.}
Since we have shown in Section \ref{subsubsection:notworking} that the CKWS-adapted refined score 
attack does provide limited scaling with having $d > 1$, we explore how well the attack performs 
assuming known-documents as the attacker's knowledge. Figure \ref{fig:varyingpknown} shows the 
accuracy of the attack using known-documents with varying known-documents rates of 
$0.05 \leq p \leq 0.8$ and steps of $0.05$. 
We observe that with the initial $|KnownQ| = 10$ setting the attack achieves higher accuracies 
faster for lower known-documents rates compared to an attack setting having $|KnownQ| = 5$ initially. 
Also, with known-documents rates $p \geq 0.7$ the accuracy of the attack becomes constant and reaches 
near $100\%$ accuracy for both $5$ and $10$ known queries. However, we do note that having a 
vocabulary size of $v_{\real} = 130$ is a rather limited setting. In the next section we 
explore the attack using known-documents with larger vocabularies.

\textbf{CKWS-adapted refined score attack using $0.7$-known-documents.}
In the previous result with varying known-documents rates we observed that the accuracy of the 
attack using known-documents reaches near $100\%$ for known-documents rate $p = 0.7$ for both 
$5$ and $10$ known queries. 
Here we explore the accuracy of the attack by fixing the known-documents rate to $p = 0.7$ with 
vocabulary sizes $250$ and $500$. Figure \ref{fig:knownbars} shows a bar plot for both these 
results with error bar describing the standard deviation of the accuracy over $50$ experiments. 
We observe that 
for vocabulary size $250$ the difference with an attack using $5$ known queries compared to $10$
known queries is small. Also, the standard deviation in both settings is small. However, for the 
$500$ keyword setting we clearly see a decrease in accuracy using $5$ known queries and a large 
standard deviation. Whereas for $10$ known queries the attack still reaches above $93\%$ accuracy 
and standard deviation is small. We do note however that in this case an attacker has great 
advantage, since it knows at least $70\%$ of the whole indexed dataset and $10$ known queries. 
In comparison, previous passive query-recovery attacks \cite{cash2015leakage,islam2012access}
on single-keyword search did not exceed $40\%$ accuracy assuming known-documents rate of $0.8$.

\begin{figure}[h!]
  \centering
  \begin{minipage}{0.49\textwidth}
    \centering
    \includegraphics[width=\textwidth]{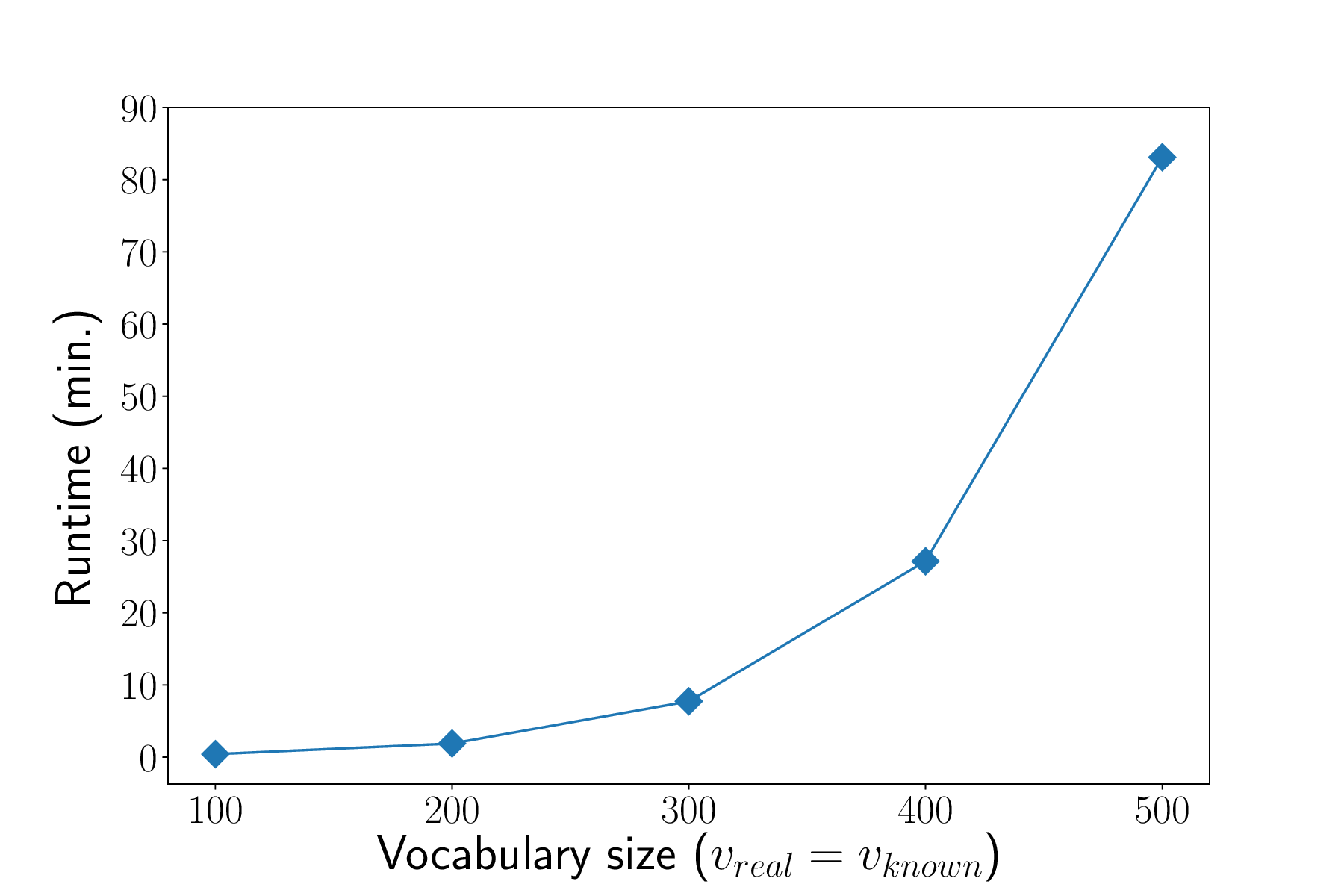}
    \caption{Runtime of the CKWS-adapted refined score attack using known-documents w.r.t.
      to vocabulary size, with $d = 2$ and $p = 0.7$.} \label{fig:runtimes}
  \end{minipage}
\end{figure}

\textbf{Runtime and memory usage.}
Figure \ref{fig:runtimes} describes the average runtime of the attack using 
known-documents over 50 repetitions in function of $v_{\real}$ for $d = 2$. 
We observe that the runtime is high for considerably small 
vocabulary sizes, which is to be expected considering the time complexity described in Section 
\ref{section:complexity}. We only show the runtime of the attack using known-documents, 
however, runtime of the attack using similar-documents is similar. 
Although our runtime can further benefit from using multiple GPUs and even our 
code is written in such fashion, we found that using two GPUs does not necessarily speed up our 
attack due to large overhead. 

The overall memory usage is dominated by the size of co-occurrence matrices $C_{\ckw}$ and 
$C_{\td}$. Therefore, we can define the main memory usage of the attack by the size of these two 
matrices as a function of the vocabulary size and the number of queries observed. In our experiments
we always assume the attacker observes $|\mathcal{Q}| = 0.15 \cdot m_{\real}$ queries. As a result 
an accurate estimation of the bytes used by one experiment is given by 
$\mathsf{numberOfBytes}(v_{\real}, d) = 2 \cdot (0.15 + 0.15^2) \cdot {v_{\real} \choose d}^2
  \cdot \mathsf{sizeof}(\mathsf{float})$, where $\mathsf{sizeof}(\cdot)$ returns the number of 
bytes used by the system to store a certain data type. Filling in for $v_{\real} = 500$, 
$d = 2$ and using $64$ bit $\mathsf{float}$, $\mathsf{numberOfBytes}(500, 2) \approx 40$ GiB, 
whereas the GPU used in our experiments fits at most $16$ GB, meaning batching intermediate 
results is already required.

\section{Discussion}

\textbf{Runtime.} Although requiring large co-occurrence matrices for the extended refined score attack
is cumbersome, if the adversary has sufficient memory resources
these large matrices will not be her only concern. Her main concern will be the runtime of the 
attack because without being able to parallelize our attack to multiple GPUs our attack is 
difficult to run for vocabulary sizes $> 500$ and becomes infeasible for vocabulary sizes $> 1000$, 
whereas the added time complexity using our extension strategy is relatively small.

\textbf{Observed queries.} Furthermore, the question arises whether it is realistic for an attacker 
to observe $15\%$ of 
all possible queries. With only single-keyword search we believe this can be achieved. However, 
with $d = 2$ the number of keyword conjunctions to be observed is big, 
i.e. $0.15 \cdot {v_{\real} \choose d}$. Although a smaller percentage could be considered more 
realistic and would even decrease the runtime of the attack, larger $|\mathcal{Q}|$ is still desired, 
since it will result in better estimators for prediction and thus higher accuracies.

\textbf{Query distribution.} In our experiments we only sampled queries 
using a uniform distribution. However, it is likely that this is unrealistic for keyword 
conjunctions, since certain keywords might be more likely to be used in a query together whereas 
other possible conjunctions might not be queried at all. Having knowledge of whether certain keywords 
are more likely to be searched for in conjunction would decrease the complexity of the attack, 
since one can then only consider the top most likely keyword conjunctions.

\textbf{Countermeasure.} Previous query-recovery attacks on single-keyword search also describe a countermeasure 
against their attack. In our work we focus on the question if a generic extension is 
possible. However, because of our generic extension strategy, countermeasures tested in 
\cite{queryvolution2020} will be applicable but were not explored. Also, 
most introduced countermeasures do not actually leak less information, they  
make the leakage unusable by the attack proposed in the corresponding work 
(e.g. adding false positives in the result set).

\textbf{Generic extension.} Although we described an adapted version of the refined score 
attack by \cite{queryvolution2020} to a conjunctive keyword setting since it is good 
performing with low runtimes for single-keywords, our generic extension strategy using 
keyword conjunction sets is also valid for other attacks 
(\cite{islam2012access,cash2015leakage,blackstone2019revisiting}) and even other types of attacks 
(e.g. attacks using query frequency \cite{liu2014search,oya2021hiding}). 
However, we expect similar runtime issues due to the large query space. 
Blackstone et al. \cite{blackstone2019revisiting} has a particular algorithm using 
cross-filtering that could be helpful to be an attack specifically against conjunctive 
keyword search.

\section{Conclusion}
In this work we presented a generic extension strategy to adapt any passive query-recovery attack 
to a conjunctive keyword search setting. We specifically explored its applicability using the 
refined score attack proposed by Damie et al. \cite{queryvolution2020} to a conjunctive keyword 
search setting. It is the first study of passive query-recovery attacks in the conjunctive keyword 
search setting. We showed that our attack using documents that are similar, but otherwise different 
from the indexed documents on the server, does only achieve accuracy of $32\%$ as attack on 
conjunctive keyword search. However, applying the adapted attack using known-documents can still 
perform with a low number of known queries and vocabulary size of $500$ and achieves a recovery 
rate similar to previous passive query-recovery attacks 
\cite{islam2012access,cash2015leakage,ning2018passive} against 
single-keyword search.

Further, we discussed that the time complexity of the adapted attack grows combinatorial with 
the number of keywords in the conjunctive search query. Also, the storage required to 
perform the attack is dominated by the size of the co-occurrence matrices computed from the 
attacker's knowledge which also increases combinatorial. 

%
%
%
\bibliographystyle{splncs04}
\bibliography{mybibliography}

\begin{thebibliography}{10}
\providecommand{\url}[1]{\texttt{#1}}
\providecommand{\urlprefix}{URL }
\providecommand{\doi}[1]{https://doi.org/#1}

\bibitem{abadi2016tensorflow}
Abadi, M., Barham, P., Chen, J., Chen, Z., Davis, A., Dean, J., Devin, M.,
  Ghemawat, S., Irving, G., Isard, M., et~al.: Tensorflow: A system for
  large-scale machine learning. In: 12th USENIX symposium on operating systems
  design and implementation ($\{$OSDI$\}$ 16) (2016)

\bibitem{blackstone2019revisiting}
Blackstone, L., Kamara, S., Moataz, T.: Revisiting leakage abuse attacks. IACR
  Cryptol. ePrint Arch.  \textbf{2019}, ~1175 (2019)

\bibitem{cash2015leakage}
Cash, D., Grubbs, P., Perry, J., Ristenpart, T.: Leakage-abuse attacks against
  searchable encryption. In: Proceedings of the 22nd ACM SIGSAC conference on
  computer and communications security (2015)

\bibitem{cash2013highly}
Cash, D., Jarecki, S., Jutla, C., Krawczyk, H., Ro{\c{s}}u, M.C., Steiner, M.:
  Highly-scalable searchable symmetric encryption with support for boolean
  queries. In: Annual cryptology conference. Springer (2013)

\bibitem{clement_2020}
Clement, J.: U.s. online search query size in 2020.
  \url{https://www.statista.com/statistics/269740/number-of-search-terms-in-internet-research-in-the-us/}
  (Aug 2020)

\bibitem{queryvolution2020}
Damie, M., Hahn, F., Peter, A.: A highly accurate query-recovery attack against
  searchable encryption using non-indexed documents. In: 30th USENIX Security
  Symposium (USENIX Security 21). USENIX Association (Aug 2021)

\bibitem{fairouz2016symmetric}
Fairouz, S.A., Lu, S.F.: Symmetric key encryption with conjunctive field free
  keyword search scheme. Journal of Advances in Mathematics and Computer
  Science  \textbf{16}(6),  1--11 (2016)

\bibitem{golle2004secure}
Golle, P., Staddon, J., Waters, B.: Secure conjunctive keyword search over
  encrypted data. In: International conference on applied cryptography and
  network security. Springer (2004)

\bibitem{hu2019forward}
Hu, C., Song, X., Liu, P., Xin, Y., Xu, Y., Duan, Y., Hao, R.: Forward secure
  conjunctive-keyword searchable encryption. IEEE Access  \textbf{7} (2019)

\bibitem{islam2012access}
Islam, M.S., Kuzu, M., Kantarcioglu, M.: Access pattern disclosure on
  searchable encryption: ramification, attack and mitigation. In: Ndss.
  vol.~20. Citeseer (2012)

\bibitem{jho2013symmetric}
Jho, N.S., Hong, D.: Symmetric searchable encryption with efficient conjunctive
  keyword search. KSII Transactions on Internet \& Information Systems
  \textbf{7}(5) (2013)

\bibitem{klimt2004introducing}
Klimt, B., Yang, Y.: Introducing the enron corpus. In: CEAS (2004)

\bibitem{lai2018result}
Lai, S., Patranabis, S., Sakzad, A., Liu, J.K., Mukhopadhyay, D., Steinfeld,
  R., Sun, S.F., Liu, D., Zuo, C.: Result pattern hiding searchable encryption
  for conjunctive queries. In: Proceedings of the 2018 ACM SIGSAC Conference on
  Computer and Communications Security (2018)

\bibitem{liu2014search}
Liu, C., Zhu, L., Wang, M., Tan, Y.A.: Search pattern leakage in searchable
  encryption: Attacks and new construction. Information Sciences  \textbf{265}
  (2014)

\bibitem{ning2018passive}
Ning, J., Xu, J., Liang, K., Zhang, F., Chang, E.C.: Passive attacks against
  searchable encryption. IEEE Transactions on Information Forensics and
  Security  \textbf{14}(3) (2018)

\bibitem{oya2021hiding}
Oya, S., Kerschbaum, F.: Hiding the access pattern is not enough: Exploiting
  search pattern leakage in searchable encryption. In: 30th USENIX Security
  Symposium (USENIX Security 21) (2021)

\bibitem{patranabis2020forward}
Patranabis, S., Mukhopadhyay, D.: Forward and backward private conjunctive
  searchable symmetric encryption. Cryptology ePrint Archive  (2020)

\bibitem{poddar2020practical}
Poddar, R., Wang, S., Lu, J., Popa, R.A.: Practical volume-based attacks on
  encrypted databases. In: 2020 IEEE European Symposium on Security and Privacy
  (EuroS\&P). IEEE (2020)

\bibitem{poon2015efficient}
Poon, H.T., Miri, A.: An efficient conjunctive keyword and phase search scheme
  for encrypted cloud storage systems. In: 2015 IEEE 8th International
  Conference on Cloud Computing. IEEE (2015)

\bibitem{porter1980algorithm}
Porter, M.F.: An algorithm for suffix stripping. Program  (1980)

\bibitem{pouliot2016shadow}
Pouliot, D., Wright, C.V.: The shadow nemesis: Inference attacks on efficiently
  deployable, efficiently searchable encryption. In: Proceedings of the 2016
  ACM SIGSAC conference on computer and communications security (2016)

\bibitem{song2000practical}
Song, D.X., Wagner, D., Perrig, A.: Practical techniques for searches on
  encrypted data. In: Proceeding 2000 IEEE Symposium on Security and Privacy.
  S\&P 2000. IEEE (2000)

\bibitem{sun2015catch}
Sun, W., Liu, X., Lou, W., Hou, Y.T., Li, H.: Catch you if you lie to me:
  Efficient verifiable conjunctive keyword search over large dynamic encrypted
  cloud data. In: 2015 IEEE Conference on Computer Communications (INFOCOM).
  IEEE (2015)

\bibitem{wang2019toward}
Wang, Y., Wang, J., Sun, S., Miao, M., Chen, X.: Toward forward secure sse
  supporting conjunctive keyword search. IEEE Access  \textbf{7} (2019)

\bibitem{wu2019vbtree}
Wu, Z., Li, K.: Vbtree: forward secure conjunctive queries over encrypted data
  for cloud computing. The VLDB Journal  \textbf{28}(1) (2019)

\bibitem{zhang2018privacy}
Zhang, L., Zhang, Y., Ma, H.: Privacy-preserving and dynamic multi-attribute
  conjunctive keyword search over encrypted cloud data. IEEE Access  \textbf{6}
  (2018)

\bibitem{zhang2016all}
Zhang, Y., Katz, J., Papamanthou, C.: All your queries are belong to us: The
  power of file-injection attacks on searchable encryption. In: 25th USENIX
  Security Symposium (USENIX Security 16) (2016)

\end{thebibliography}

\end{document}